\documentclass[10pt]{article}
\usepackage{float}
\usepackage[symbol]{footmisc}
\usepackage[superscript,sort]{cite}
\usepackage{array}
\usepackage[normalem]{ulem} 
\usepackage{amsmath,amssymb}

\usepackage{amsthm,amscd,amsxtra,amsfonts,amsmath,amssymb,multirow}
\usepackage{wrapfig}
\usepackage[footnotesize]{caption}
\usepackage{subcaption}
\usepackage[tiny,compact]{titlesec}
\usepackage{booktabs}
\usepackage{siunitx}
\usepackage{threeparttable} 
\usepackage{helvet}

\usepackage{xcolor,colortbl}
\usepackage{tikz}
\usetikzlibrary{shapes,arrows}
\usepackage[T1]{fontenc}
\usepackage[linesnumbered,ruled]{algorithm2e} 
\titlespacing*{\section}{0pt}{*0}{*0}
\titlespacing*{\subsection}{0pt}{*0}{*0}
\titlespacing*{\subsubsection}{0pt}{*0}{*0} 
\titlespacing{\paragraph}{0pt}{*0}{*1}

\definecolor{MyPurple}{rgb}{1,0,1}

\newcommand{\barray}{\begin{array}{ll}}
\newcommand{\earray}{\end{array}}

\usepackage{mathtools}
\begin{document}
\pagenumbering{roman}

\clearpage \pagebreak \setcounter{page}{1}
\renewcommand{\thepage}{{\arabic{page}}}

\title{Mathematical deep learning for pose and binding affinity prediction and ranking in D3R Grand Challenges
}

\author{Duc Duy Nguyen$^{1}$,  Zixuan Cang$^{1}$, Kedi Wu$^{1}$, Menglun Wang$^{1}$, Yin Cao$^{1}$ and  Guo-Wei Wei$^{1,2,3,}$\footnote{
		Corresponding to Guo-Wei Wei.		Email: wei@math.msu.edu}\\
	$^1$ Department of Mathematics,
	Michigan State University, MI 48824, USA.\\
	$^2$ Department of Electrical and Computer Engineering,
	Michigan State University, MI 48824, USA. \\
	$^3$ Department of Biochemistry and Molecular Biology,
	Michigan State University, MI 48824, USA. \\
}

\date{\today}
\maketitle

\begin{abstract}
Advanced mathematics, such as multiscale weighted colored graph and element specific persistent homology, and machine learning including deep neural networks  were integrated to construct mathematical deep learning models
for pose and binding affinity prediction and ranking in the last two D3R grand challenges in computer-aided drug design and discovery.  D3R Grand Challenge 2 (GC2) focused on the pose prediction and binding affinity ranking and free energy prediction for Farnesoid X receptor ligands. Our models obtained the top place in absolute free energy prediction for free energy Set 1 in Stage 2.  The latest competition, D3R Grand Challenge 3 (GC3),  is  considered as the most difficult challenge so far. It has 5 subchallenges involving Cathepsin S and five other kinase targets, namely VEGFR2, JAK2, p38-$\alpha$, TIE2, and ABL1. There is a total of 26 official competitive  tasks for GC3. Our predictions were ranked 1st in 10 out of  26 official competitive  tasks.

\end{abstract}
\maketitle

\pagestyle{empty}

\newpage
{\setcounter{tocdepth}{4} \tableofcontents  }

\newpage

\section{Introduction}
With the  availability of increasingly powerful computers and fast accumulating molecular and biomolecular datasets,  one can dream of a possible scenario that all the major tasks of drug design and discovery can be conducted on computers  \cite{Berman:2000, liu2017forging, ahmed2014recent}. Virtual screening (VS) is one of the most important aspects of computer-aid drug design (CADD) \cite{kroemer2007structure}. VS involves two stages, namely, the generation of  different ligand conformations (i.e., poses) when a compound is docked to a target protein binding site, and  the  prediction of   binding affinities. It is generally believed that   the first stage can be well resolved by available techniques, such as  molecular dynamics (MD), Monte Carlo (MC), and genetic algorithm (GA)  \cite{Leach:2006,Novikov:2011,RenxiaoWang:2003CompareSF}. However, The development of scoring function (SF) for binding affinity prediction  with high accuracy still remains a formidable challenge. In general, current SFs can be classified into four different categories, namely force-field-based ones, knowledge-based ones, empirical-based ones and machine learning-based ones \cite{LiuJie:2014}. Among them,  force-field-based SFs, such as  COMBINE \cite{Ortiz:1995} and MedusaScore  \cite{Yin:2008}, emphasize the physical description of protein and ligand interactions in the solvent environment, including  van der Walls (vdW), electrostatics, hydrogen bonding, solvation effect, etc. Typical Knowledge-based SFs represent the binding affinity as the linear sum of pairwise statistical potentials between receptor and ligand atoms.  KECSA \cite{Merz:2015Solvation}, PMF \cite{PMFScore:1999}, DrugScore \cite{DrugScore:2005}, and IT-Score \cite{ITScore:2006} are  some of the well-known examples. The empirical-based SFs, in fact, make use of multiple linear regression to construct a linear combination from different physical-descriptor components such as vdW interaction, hydrophobic, hydrogen bonding, desolvation, dipole, etc. The renowned candidates for empirical-based SFs include X-Score \cite{XScore:2002}, PLP \cite{Verkhivker:1995PLP}, and ChemScore \cite{Eldridge:1997}, etc.

Recently, machine learning including deep learning has  emerged as a major technique in CADD. By using advanced machine learning algorithms, such as random forest (RF) and deep convolutional neural network, the machine learning-based  SFs can characterize the non-additive contributions of functional groups in  protein-ligand binding interactions \cite{Baum:2010}. Such a characterization can help machine learning-based SFs consistently maintain their accuracy in binding affinity predictions for a variety of protein-ligand complexes \cite{li:2014,DDNguyen:2017d,ZXCang:2017b,ZXCang:2017c,ZXCang:2018a}. However, the performance of machine learning-based SFs depends crucially on the training data quality and statistic distribution. Additionally, it also depends on selected features that might or might not accurately and completely describe the protein-ligand binding interactions. We assume that the physics of interest of complex biomolecules and  interactions lies on low-dimensional manifolds or subspaces embedded in a high-dimensional data space. Based on this hypothesis, we have recently proposed several low-dimensional mathematical models that dramatically reduce the structural complexity of protein-ligand complexes and give rise to surprisingly accurate predictions of various  bimolecular properties. For example,  we proposed a multiscale weighted colored graph (MWCG) model to simplify   protein structures and analyze their  flexibility  \cite{bramer2018multiscale}. The essential idea of this method is to use the graph theory to represent the interactions between atoms in a molecule in an element-level collective manner.  The WMCG approach has been shown to be over 40\% more accurate than the Gaussian network model  on a set of 364 proteins \cite{bramer2018multiscale}. 

In addition to graph theory simplification, we have also developed the topological abstraction of complex protein structures. In order to describe  the topological changes such as the opening or closing of ion channels, the folding or unfolding of proteins, and the subtle change in binding site after the protein-ligand binding, we take the advantage of topological methods to study the connectivity of different molecular components in a space \cite{Kaczynski:2004} which can represent the independent topological entities such as independent components, rings and higher dimension faces. However, since the conventional topology and homology  are metric or coordinate free, they capture very little biomolecular geometric information and thus are unable to efficiently characterize biomolecular structures. Persistent homology (PH) is a new branch in algebraic topology. It embeds the geometric information into topological invariants. By changing a filtration parameter such as the radius of atoms PH creates a family of topological structures for a given set of atoms. As a result, the topological properties of a given biomolecule can be systematically analyzed and recorded in terms of topological invariants, i.e., the so-called Betti numbers, over the filtration process. The resulting barcodes monitor the ``birth'' and ``death'' of isolated components, circles, and cavities at different geometric scales. The persistent homology framework together with practical algorithms was introduced by Edelsbrunner \textit{et al.} \cite{Edelsbrunner01topologicalpersistence} and formal mathematical theories were also developed by Zomorodian and Carlsson \cite{Zomorodian:2005}. A zeroth dimensional version was also introduced earlier under the name of size function by Frosini and Landi \cite{Frosini:1999}. Primitive applications of PH to computational biology has been reported in the literature  \cite{Kasson:2007,Gameiro:2014,dabaghian2012topological}. Recently, we have developed a variety of advanced PH models to analyze the topology-function relationship in protein folding and protein flexibility \cite{KLXia:2014c},  quantitative predictions of curvature energies of fullerene isomers \cite{KLXia:2015a,BaoWang:2016a}, protein cavity detection \cite{ESES:2017}, and the resolving  ill-posed inverse problems in cryo-EM structure determination \cite{KLXia:2015b}. In 2015, we introduced some of the first combinations of PH and machine learning for  protein structural classification  \cite{ZXCang:2015}. Topological descriptors were further integrated with  a variety of deep learning algorithms to achieve state-of-the-art analysis and prediction of protein folding stability change upon mutation \cite{ZXCang:2017a}, drug toxicity   \cite{KDWu:2018a}, aqueous solubility, partition coefficient,  \cite{wu2017topp}, binding affinity \cite{ZXCang:2017b,ZXCang:2017c} and virtual screening \cite{ZXCang:2018a}.

In this paper, we report the performance of our  mathematical deep learning models on pose and binding affinity prediction and ranking in the last two D3R grand challenges, namely D3R Grand Challenge 2 (GC2) and D3R Grand Challenge 3 (GC3). The GC2 was initiated in 2016 and consisted of two stages. The first stage asked participants to predict the crystallographic poses of 36 ligands for the target of farnesoid X receptor (FXR). In addition, there were affinity ranking task for all 102 compounds and absolute free energy prediction for two designated subsets of 18 and 15 small molecules. In the second stage, participants were asked again to submit the affinity ranking and free energy after the release of 36 crystal structures. In GC2, we employed our mathematical deep learning  models to select the best poses from  docking software generated poses for  binding affinity ranking and prediction tasks. Our models achieved the top place in affinity ranking for the free energy Set 1 in Stage 2.

In addition, our  results for the latest grand challenge, i.e., GC3, are presented in this paper. The third grand challenge, took place in 2017, is the largest in terms of the number of competitive tasks since 2015. It consisted of 5 subchallenges. Subchallenge 1 was about  Cathepsin S. It comprised two stages with  tasks the same as ones in the GC2. There were 24 ligands with crystal structures and their binding energies spread  three orders of magnitude for 136 compounds. Subchallenge 2 focused on  kinase selectivity. It has three kinase targets, namely VEGFGR2, JAK2, and p38-$\alpha$ with their numbers of compounds being 85, 89, and 72, respectively. This subchallenge only asked participants to submit affinity ranking for each kinase dataset. Subchallenge 3 involved the binding affinity ranking and free energy prediction of target JAK2. It consisted of a relatively small dataset with 17 ligands having similar chemical structures. In Subchallenge 4, there were 18 congeneric ligands with Kd values for  kinase TIE2. In addition to asking for the affinity ranking of 18 compounds, Subchallenge 4 asked participants to predict the free energies of two subsets with 4 and 6 compounds, respectively. The last subchallenge in  GC3 concerned the binding affinity ranking of different mutants on protein ABL1. There were two compounds and 5 different mutation sites. Overall, our models performed well in GC3.  Specifically, we obtained the first place in 10 out of a total of 26 predictive tasks.

\section{Methods}
In this section, we briefly describe our computation methods and algorithms developed for GC2 and GC3.

\subsection{Ligand preparation}
All ligands in  Grand Challenges are provided in the SMILES string format. They are converted to the optimal 3D structures and protonated at pH=7.5 using \texttt{ligprep} tool in Schr\"{o}dinger software \cite{Madhavi:2013Schrodinger}. Before employing Autodock Vina \cite{Trott:2010AutoDock} for docking, Gasteiger partial charges were added to these ligands via MGLTools v1.5.6  \cite{morris2009autodock4}.

\subsection{Protein structures selection and preparation}
Except for Subchallenge 1, all the receptor structures in GC3 are supplied in the protein sequence format. We utilized the homology modeling task in Maestro of Schr\"{o}dinger software \cite{bell2012primex} to obtain  3D structure predictions. In addition, we make use of the crystal structures available in the Protein Data Bank (PDB) for each protein family (see the supporting information for a complete list). These collected protein structures were prepared using the protein preparation wizard provided in Schr\"{o}dinger package  \cite{Madhavi:2013Schrodinger} with default parameters except enabling the \texttt{fillsidechains} option.

\subsection{Docking protocols}
We use a number of docking protocols in GC2 and GC3. Among, a machine learning protocol was developed  in our own lab.  Motivated by  earlier  work  \cite{ye2016optimal}, we carried out four different docking strategies, namely align-close, align-target, close-dock and cross-dock, to attain the best poses for  binding affinity predictions. We also used  induced fit docking (IFD) and unrestricted IFD in our pose predictions. 

\paragraph{Protocol 1: Machine learning based docking} We developed a machine learning-based scoring function to select the poses generated by GOLD \cite{G-Score}, GLIDE \cite{Friesner:2004Schrodinger}, and Autodock Vina \cite{Trott:2010AutoDock}. Given a ligand target, we at first formed a training data of complexes taken from the PDB. The criteria for such  selections are based on the similarity coefficient, measured by fingerprint 2 (FP2) in Open Babel v2.3.1 \cite{o:2011}, of ligand in the complex. Then, we utilized docking software packages such as GOLD, GLIDE, and Autodock Vina to re-dock ligands to protein in those selected complexes. A variety of docking poses was distributed into 10 different RMSD bins as follows: [0,1], (1,2], \dots, (9,10]\AA. In each bin, we clustered decoys into 10 clusters based on their internal similarities. The docking poses having the smallest free energy were selected as the candidate for their clusters. As a result, one may end up with a total of 100 poses for each given complex. We employed all these decoy poses to form a training set with labels defined by their RMSDs. Our topological based deep learning models were utilized to learn this training set. Finally, we employed this established scoring function to re-rank the poses of the target ligand produced by docking software packages.


\paragraph{Protocol 2: Align-close} In the align-close method, we select ligand available in the PDB that has the highest chemical similarity to the target ligand. Here, the similarity score was measured by fingerprint 2 (FP2) in Open Babel v2.3.1 \cite{o:2011}. It is also noted that all the processed structures in this procedure were conducted in the Schr\"{o}dinger suite 2017-4 \cite{schrodinger2017-4}. A ligand was aligned to its similar candidates by the flexible ligand alignment task in Schr\"{o}dinger's Maestro  \cite{dixon2006phase, dixon2006phase2}. Then, the resulting aligned ligand is minimized to the co-crystal structure of the most similar ligand by Prime in Sch\"{o}dinger package  \cite{jacobson2004hierarchical, jacobson2002role, schrodinger2017-4}. 

\paragraph{Protocol 3: Align-target}
In the align-target protocol, the homology modeling tool in Maestro was used to construct  protein 3D structures from given sequences, and the aligned ligands obtained from the align-close procedure are minimized with respect to  corresponding receptors.

\paragraph{Protocol 4: Close-dock}
The fourth docking strategy is called as close-dock. Based on previous docking methods, one can identify the most similar structure in the PDB to a given D3R ligand. This procedure also gives us the corresponding co-crystal structure, i.e.,  the so-called closet receptor. In the close-dock approach, Autodock Vina is used to docking the target ligand to its corresponding closet receptor. The best pose is selected based on  Autodock Vina's energy scoring.

\paragraph{Protocol 5: Cross-dock}
The next approach in our docking methods is named cross-dock. This is basically a cross docking method in which the close receptors are the co-crystal structures of the ligands having the similar chemical characteristics to the interested ligand. In the cross-docking procedure, we use Autodock Vina to dock the D3R ligands to their close receptors. Those poses that have the smallest binding energies are selected as the best poses.

\paragraph{Protocol 6: Constraint-IFD}
Similarly to the align-target protocol, we used the homology modeling module in Maestro to  generate 3D structure from a given sequence. For the docking procedure, we employed the induced fit docking (IFD) \cite{farid2006new, sherman2006novel, sherman2006use} in Maestro with restricting docking poses to the closet ligands with a tolerance of  3\AA. The best pose was selected due to the ranking from IFD.

\paragraph{Protocol 7: Free-IFD}
This protocol is exactly the same protocol as Constraint-IFD except for no constraint during the run of induced-fit docking.

\subsection{Multiscale weighted colored graph representation }\label{sec:mwcg}

Weighted colored graph (WCG) method describes  intermolecular and intramolecular interactions as pairwise atomic correlations \cite{bramer2018multiscale}. To apply the WCG for analyzing the protein-ligand interactions, we convert all the atoms and their pairwise interactions at the binding site of a protein-ligand complex with a cutoff distance  $d$ into a colored graph $G(V^{d}, E)$ with vertices $V^d$ and edge $E$. As such, the $i$th atom is labeled by its position ${\bf r}_i$ , element type $\alpha_i$ and co-crystal type $\beta_i$. Thus, we can express vertices $V^{d}$ as
\begin{align}
V^d=\{({\bf r}_i,\alpha_i,\beta_i)| {\bf r}_i\in \mathbb{R}^3, \alpha_i\in \mathcal{C},\beta_i\in\mathcal{S}, \|{\bf r}_i - {\bf r}_j\| < d \text{~for some $1\leq j\leq N$ such that $\beta_i+\beta_j = 1$} ,i=1,2,\dots,N\},
\end{align}
  where $\mathcal{C}=\{$C, N, O, S, P, F, Cl, Br, I$\}$ contains all the commonly occurring element types in a complex, and $\mathcal{S}=\{$0, 1$\}$ meaning that if the $i$th atom belongs to protein then $\beta_i=0$, otherwise $\beta_i=1$. Hydrogen element is omitted since it does not present in the crystal structures of most protein-ligand complexes. To describe  pairwise interactions between the protein and the ligand, 
	 we define an ordered and colored set  $\mathcal{P}=\{(\alpha,0)(\alpha',1)\}$. Here, $\alpha\in\{\text{C, N, O, S}\}$ is a heavy atom in the protein, and $\alpha'\in\{\text{C, N, O, S, P, F, Cl, Br, I}\}$ is a heavy atom in the ligand. With that setting, it is trivial to verify that set $\mathcal{P}$ has a total 36 partitions. For example, a partition $\mathcal{P}_1 = \{(C,0)(O,1)\}$ contains all pairs $CO$ in the complex with the first atom is a carbon in the protein and the second atom is an oxygen in the ligand. For each set of atom pairs $\mathcal{P}_k$, $k=1,2,\dots,36$, a set of vertices $V_{\mathcal{P}_k}$ is a subset of $V^{d}$ containing all atoms that belong to a pair in $\mathcal{P}_k$. Therefore, the edges in such WCG describing potential pairwise atomic interactions are defined by 
  \begin{align}
  E_{\mathcal{P}_k}^{\sigma,\tau,\zeta} = \{\Phi^\sigma_{\tau,\zeta}(\|{\bf r}_i - {\bf r}_j\|)|((\alpha_i,\beta_i)(\alpha_j,\beta_j))\in\mathcal{P}_k;i,j=1,2,\dots,N\},
  \end{align}
where $\|{\bf r}_i - {\bf r}_j\|$ defines a Euclidean distance between $i^{th}$ and $j^{th}$ atoms, $\sigma$ indicates the type of radial basic functions (e.g., $\sigma=\text{L}$ for Lorentz kernel, $\sigma=\text{E}$ for exponential kernel), $\tau$ is a scale distance factor between two atoms, and $\zeta$ is a parameter of power in the kernel (i.e., $\zeta=\kappa$ when $\sigma=\text{E}$, $\zeta=\nu$ when $\sigma=\text{L}$). The kernel  $\Phi^\sigma_{\tau,\zeta}$ characterizes a pairwise correlation satisfying the following conditions
\begin{align}
\Phi^\sigma_{\tau,\zeta}(\|\mathbf{r}_i-\mathbf{r}_j\|)=1 &\mbox{ as } \|\mathbf{r}_i-\mathbf{r}_j\|\rightarrow 0,\\
\Phi^\sigma_{\tau,\zeta}(\|\mathbf{r}_i-\mathbf{r}_j\|)=0 &\mbox{ as } \|\mathbf{r}_i-\mathbf{r}_j\|\rightarrow\infty.
\end{align}

Commonly used radial basis functions include generalized exponential functions
\begin{align}
\Phi^{\rm E}_{\tau,\kappa} = e^{-(\|\mathbf{r}_i-\mathbf{r}_j\|/\tau(r_i+r_j))^{\kappa}},\quad \kappa>0, 
\end{align}
and generalized Lorentz functions
\begin{align}
\Phi^{\rm L}_{\tau,\nu}(\|\mathbf{r}_i-\mathbf{r}_j\|)= \dfrac{1}{1+(\|\mathbf{r}_i-\mathbf{r}_j\|/\tau(r_i+r_j))^{\nu}},\quad \nu>0, 
\end{align}
where $r_i$ and $r_j$ are, respectively, the van der Waals radius of the $i^{th}$ and $j^{th}$ atoms.  

In the graph theory or network analysis, centrality is widely used to identify the most important nodes \cite{borgatti2005centrality}. There are various types of centrality such as degree centrality  \cite{freeman1978centrality}, closeness centrality \cite{bavelas1950communication}, harmonic centrality \cite{dekker2005conceptual}, etc. Specifically, while the degree centrality is measured as a number of edges upon a node, closeness and harmonic centralities depend on the length of edges and are defined as 
$1/\sum_j \|\mathbf{r}_i-\mathbf{r}_j\|$  and  $\sum_j 1/\|\mathbf{r}_i-\mathbf{r}_j\|$, respectively.  Our centrality used in the current work is an extension of the harmonic formulation by our correlation functions
\begin{equation}
\mu^{k,\sigma,\tau,\nu}_i=\sum_{j=1}^{|V_{{\cal P}_k|}}w_{ij}\Phi^{\sigma}_{\tau,\nu}(\|\mathbf{r}_i-\mathbf{r}_j\|), 
\quad((\alpha_i,\beta_i)(\alpha_j,\beta_j))\in \mathcal{P}_k,
\quad \forall i=1,2,\ldots, |V_{{\cal P}_k}|,
\end{equation} 
where $w_{ij}$ is a weight function assigned to each atomic pair. In the current work, we choose $w_{ij}=1$ if $\beta_i + \beta_j=1$, otherwise $w_{ij}=0$, for all calculations to reduce dimension of the parameter space. To describe a centrality for the whole graph $G(V_{\mathcal{P}_k},E_{\mathcal{P}_k}^{\sigma,\tau,\zeta})$, we take into account a summation of the node's centralities
\begin{align}
\mu^{k,\sigma,\tau,\nu} = \sum_{j=1}^{|V_{{\cal P}_k|}}\mu^{k,\sigma,\tau,\nu}_i.
\end{align}

Since we have 36 choices of the set of weighted colored edges ${\cal P}_k$, we can obtain corresponding 36 graph centralities $\mu^{k,\sigma,\tau,\nu}$. By varying kernel parameters $(\sigma,\tau,\nu)$, one can achieve multiscale centralities for multiscale weighted colored graph (MWCG) \cite{bramer2018multiscale}. For a two-scale WCG, we obtain a total of 72 descriptors for a protein-ligand complex. 


\subsection{Algebraic topology based molecular signature}

The geometry of biomolecular systems together with the complex interaction patterns allows us to build topological spaces upon the systems which facilitate  powerful topological analysis. The topological analysis provides us a description of the molecular system that captures a collection of key aspects of the system including the multiscale description of geometry, the characterization of interaction network in an arbitrary dimension, and the important physical and chemical information, which ensures the success of the downstream machine learning modeling. In this section, we first briefly describe the background of persistent homology. Then, we demonstrate how to  apply it to biomolecular systems to obtain a rich but concise description.

\subsubsection{Persistent homology}

We describe the theory of persistent homology in the framework of simplicial homology in a geometric sense where topological spaces are represented by collections of points, edges, triangles, and their higher-dimensional counterparts. A $k$-simplex is a collection of $(k+1)$ affinely independent points in $\mathbb{R}^n$ with $n\geq k$. If the vertices of a simplex is a subset of the vertices of another simplex, it is called a face of the other simplex. Simplices of various dimensions are building blocks of a simplicial complex which is a finite collection of simplices satisfying two conditions: (1) the faces of any simplex in the complex are also in the complex and (2) the intersection of two simplices in the complex is either empty or a common face of the two. A simplicial complex can be used to discretely represent or approximate a topological space. Given a simplicial complex $X$, a $k$-chain is a formal sum of all the $k$-simplices in $X$ which is defined as
\begin{equation}
c = \sum\limits_i a_i\sigma_i,
\end{equation}
where $\sigma_i$ is a $k$-simplex in $X$ and $a_i$ is a coefficient in a coefficient set of choice such as a finite field $\mathbb{Z}_p$ with a prime $p$. The set of all $k$-chains with the addition operator in the coefficient group forms a group called the $k$th chain group denoted $\mathcal{C}_k(X)$. The chain groups of different dimensions are connected by a collection of homeomorphisms called the boundary operators forming a chain complex,
\begin{equation}
	\cdots
	\xrightarrow{\mathmakebox[.5cm]{\partial_{i+1}}} \mathcal{C}_{i}(X)
	\xrightarrow{\mathmakebox[.5cm]{\partial_{i}}} \mathcal{C}_{i-1}(X)
	\xrightarrow{\mathmakebox[.5cm]{\partial_{i-1}}}
	\cdots
	\xrightarrow{\mathmakebox[.5cm]{\partial_{2}}} \mathcal{C}_{1}(X)
	\xrightarrow{\mathmakebox[.5cm]{\partial_{1}}} \mathcal{C}_{0}(X)
	\xrightarrow{\mathmakebox[.5cm]{\partial_{0}}} 0.
\end{equation}
It suffices to define the boundary operator on simplices, and then, such a definition can  be extended to general chains.
\begin{equation}
\partial_k(\sigma) = \sum\limits_{i=0}^k(-1)^i[v_0,\cdots,\hat{v}_i,\cdots,v_k],
\end{equation}
where $v_0,\cdots,v_k$ are vertices of the $k$-simplex $\sigma$ and $[v_0,\cdots,\hat{v}_i,\cdots,v_k]$ means the codim-$1$ face of $\sigma$ be omitting the vertex $v_i$. The boundary operator has an important property that 
\begin{equation}\label{eq:bmprpty}
\partial_k\circ\partial_{k+1}=0.
\end{equation}
With the boundary operators, we can define boundary groups and cycle groups which are subgroups of chain groups. The $k$th boundary group is defined to be the image of $\partial_{k+1}$ denoted $\mathcal{B}_k(X)=\mathrm{Im}(\partial_{k+1})$. The $k$th cycle group is defined to be the kernel of $\partial_k$ denoted $\mathcal{Z}_k(X)=\mathrm{Ker}(\partial_k)$. It can be seen that $\mathcal{B}_k(X)\subseteq \mathcal{Z}_k(X)$ following the property in Eq.~(\ref{eq:bmprpty}). Then, the $k$th homology group is defined to be the quotient group
\begin{equation}
\mathcal{H}_k(X)=\mathcal{Z}_k(X)/\mathcal{B}_k(X).
\end{equation}
Intuitively, the $k$th homology group contains elements associated to $k$ dimensional holes which are not boundaries of $(k+1)$-chains to characterize the topology.

The theory described above computes the homology of various dimensions of a topological space to obtain a multidimensional characterization of the space. However, this is not enough for the cases where the objects are also multiscale. Therefore, instead of only computing homology for a fixed topological space, we can build a sequence of subspaces of the topological space and track how homology evolves along this changing sequence. This sequence is called a filtration,
\begin{equation}
\emptyset = X_0\subseteq X_1\subseteq \cdots\subseteq X_{m-1}\subseteq X_m=X.
\end{equation}
The filtration naturally induces an inclusion map connecting the homology groups of a certain dimension,
\begin{equation}
\mathcal{H}_k(X_0)\rightarrow\mathcal{H}_k(X_1)\rightarrow\cdots
\rightarrow\mathcal{H}_k(X_{m-1})\rightarrow\mathcal{H}_k(X_m).
\end{equation}
Then for a homology generator $\delta\in\mathcal{H}_k(X_i)$, it is said to be born at $i$ if it is not an image of the inclusion map from $\mathcal{H}_k(X_{i-1})$ and it is said to die at $i+1$ if it mapped to the empty set or another homology generator that is born before $i$ by the inclusion map from $\mathcal{H}_k(X_i)$. Persistent homology tracks how these homology generators appear and disappear along the course of the filtration resulting in a robust multiscale description of the original topological space. The birth and death of each generator can be represented by a half-open interval starting at the birth time and stopping at the death time. There are several visualization methods for collections of such intervals such as barcodes and persistence diagrams.

\subsubsection{Topological description of molecular systems}

To describe molecular systems using persistent homology, the atoms can be regarded as vertices and different constructions of filtrations can reveal different aspects of the system.

To describe a complex protein geometry, an efficient filtration using alpha complex \cite{Alpha} can be employed. To build an alpha filtration, a Voronoi diagram is first generated for the collection of points representing the atoms in the system. The final frame of the topological spaces at the end of the course of filtration is constructed by including a $k$-simplex if there is a nonempty intersection of the $(k+1)$ Voronoi cells associated to its $(k+1)$ vertices. The filtration of the space can be constructed by associating a subcomplex to each value of a filtration parameter $\epsilon$. The subcomplex associated to $\epsilon$ is defined as
\begin{equation}
X_{alpha}(\epsilon)=\{\sigma\in X| \sigma=[v_0,\cdots,v_k],\, \cap_i\left(V(v_i)\cap B_\epsilon(v_i)\right)\neq \emptyset \},
\end{equation} 
where $V(v_i)$ is the Voronoi cell of $v_i$ and $B_\epsilon(v_i)$ is an $\epsilon$ ball centered at $v_i$.

A more abstract construction of filtration using the Vietoris-Rips complex can be used to address other properties of the system such as protein-ligand interactions. Given a set of points with a pairwise distance (not necessarily satisfying triangular inequality), the subcomplex associated to a filtration parameter $\epsilon$ is defined to be
\begin{equation}
X_{rips}(\epsilon) = \{\sigma\in X| \sigma=[v_0,\cdots,v_k],\, d(v_i,v_j) \leq 2\epsilon\,\mathrm{for}\,0\leq i,j\leq k \},
\end{equation}
where $d$ is the predefined distance function and $X$ is the collection of all possible simplices. Tweaking the distance function can help emphasize on different properties of the system. For example, in a protein-ligand complex, setting the distance between an atom from the protein and an atom from the ligand to the Euclidean distance while setting the distance between atoms from the same molecule to infinity will emphasize the interaction pattern between the two molecules \cite{ZXCang:2017b}. Also, we can assign values between atoms according to a specific distance of interest by using kernel functions as distances \cite{ZXCang:2017b}. We have proposed element specific persistent homology to encode physical interactions into topological invariants \cite{ZXCang:2017a,ZXCang:2017b}. By computing persistent homology on subsets of the atoms, we can extract different chemical information. For example, the element specific  persistent homology computation on the collection of all carbon atoms describes the hydrophobic network or the structural basis of the molecule while computation on the nitrogen and oxygen atoms characterizes   the hydrophilic networks \cite{ZXCang:2017b}. For the characterization of small molecules, we can use a multilevel element specific persistent homology to both describe the covalent bonds and noncovalent interactions in the molecule \cite{ZXCang:2018a}.

The element specific persistent homology  results (barcodes) can be paired with machine learning models in several ways. For example, Wasserstein metric can be applied to measure similarities among the barcodes of different proteins, which can be used with methods such as nearest neighbors and manifold learning \cite{ZXCang:2018a}. The element specific persistent homology barcodes can also be turned into fixed length feature vectors by discretizing the range of barcode and counting the persistence, birth, and death events that fall in each subinterval. The statistics of    element specific persistent homology barcodes can also be used for featurization \cite{ZXCang:2018a}. These fixed length features can be used with powerful machine learning methods such as the ensemble of trees and deep learning neural networks \cite{ZXCang:2017b,ZXCang:2017c}. The barcodes can also be transformed to representations similar to images and used in a 1-dimensional or a 2-dimensional convolutional neural networks \cite{ZXCang:2017c,ZXCang:2018a}.

\subsection{Machine learning algorithms}

\begin{figure}[!ht]
	\centering
	\includegraphics[width=0.8\textwidth]{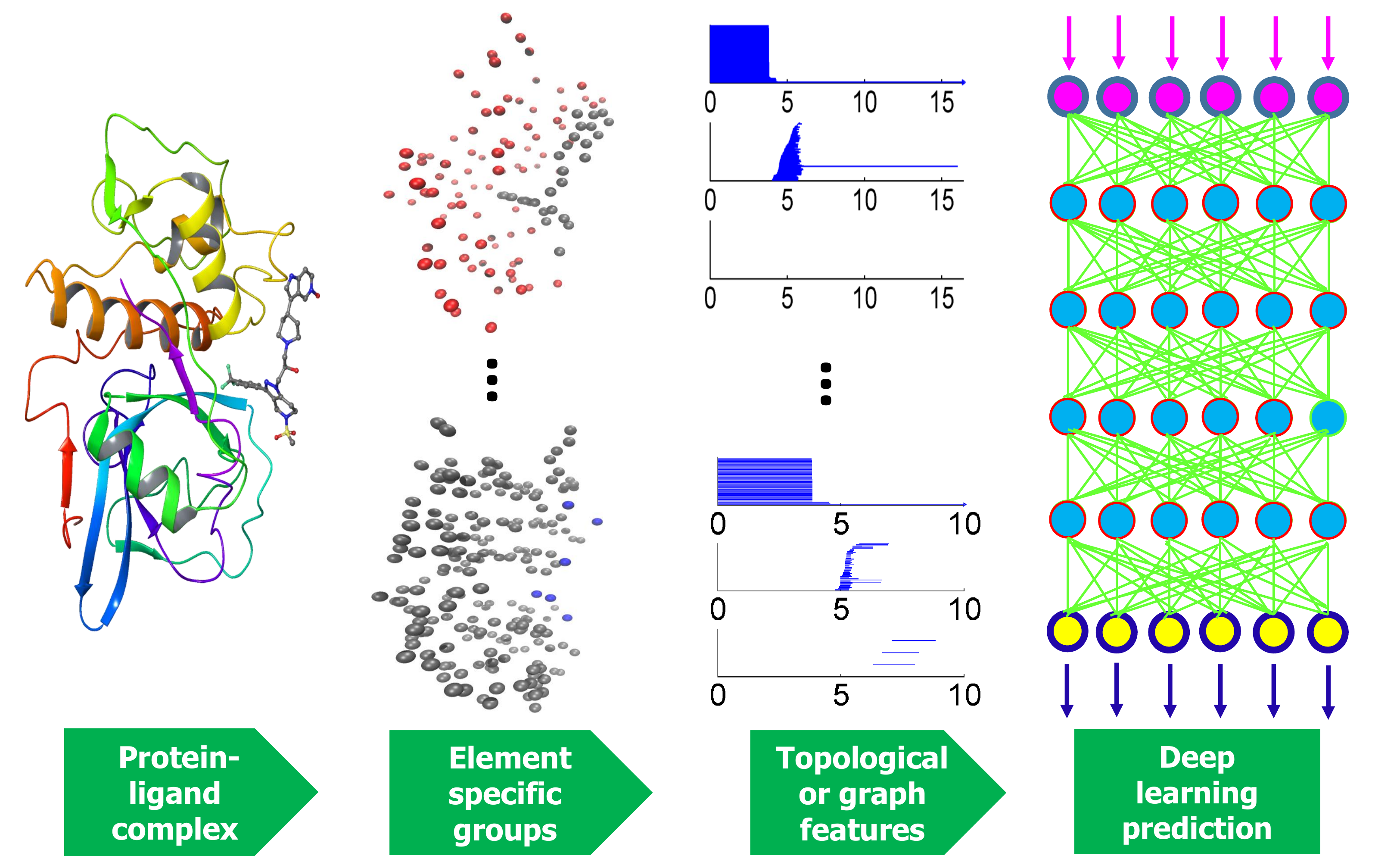}
	\caption{Illustration of mathematical deep learning prediction.}
	\label{fig:flowchart}
\end{figure}
The machine learning methods used in our prediction fall into two categories, ensemble of trees and deep learning. A schematic illustration of our mathematical deep learning modeling  is given in Fig. \ref{fig:flowchart}.

\paragraph{Ensemble of trees}
The basic building block of this type of method is decision tree which identifies key features to make decisions at the nodes of the tree. Due to its simple structure, it is usually considered as weak learner especially in the case of highly nonlinear problems or applications with high dimensional features. Ensemble of trees methods build models consisting of a collection of decision trees with the assumption that grouping the weak learners can improve the learning capability. We mainly used random forest and gradient boosting trees for the prediction. Random forest builds uncorrelated decision trees with each tree being trained on a resampling of the original training set (bootstrap). On the contrary, gradient boosting trees add one tree to the collection at a time along the direction of the steepest descent of the loss of the current collection. As these two models attempt to reduce error in two different ways, they behave differently in the bias-variance tradeoff where the random forest is better at lowering bias and gradient boosting trees focus more on reducing variance. Therefore, a higher level bagging of models of different kinds can further improve the performance. The ensemble learning methods are also robust and overfitting can be reduced by learning partial problems. For example, each tree can be trained with a random subset of the training data and a subset of the features and the model complexity can be constrained by setting maximum tree depth. Both our graph theory based models \cite{DDNguyen:2017d,DDNguyen:2018b} and algebraic topology based models \cite{ZXCang:2017b,ZXCang:2018a} achieve top-class performance with the ensemble of trees methods. 

\paragraph{Deep Learning}
When the feature is complex or there is some underlying dimension in the feature space, deep learning models can further push the performance of the predictor. For example, a spatial dimension associated to the filtration parameter lies in the persistent homology representation of  protein-ligand systems. This enables the usage of the powerful convolutional neural networks (CCNs) which has been extremely successful in the field of computer vision. The neural networks we used in the prediction are in the category of feedforward networks where the signal from the previous layer undergoes a linear transformation to the current layer, then the current layer applies a nonlinear activation function and sends the signal to the next layer. Classical deep neural networks are constructed by stacking fully connected layers where every pair of neurons in two adjacent layers are connected. Different rules of neuron connections and parameter sharing have resulted in a number of powerful deep learning models that flourish in various application fields. CNNs take advantage of the feature structure where there are spatial dimensions and only allow local connections with the parameters shared along the spatial dimensions which significantly lowers the dimension of the parameter space. Also, the flexibility of neural networks allows learning different but related tasks together by sharing layers, i.e., a type of multi-task learning. We applied convolutional neural networks and multi-task learning in our prediction which further advanced the capability of our models \cite{ZXCang:2017c,ZXCang:2018a}.

\section{Results and discussion}

\begin{figure}[!ht]
	\centering
	\includegraphics[width=0.8\textwidth]{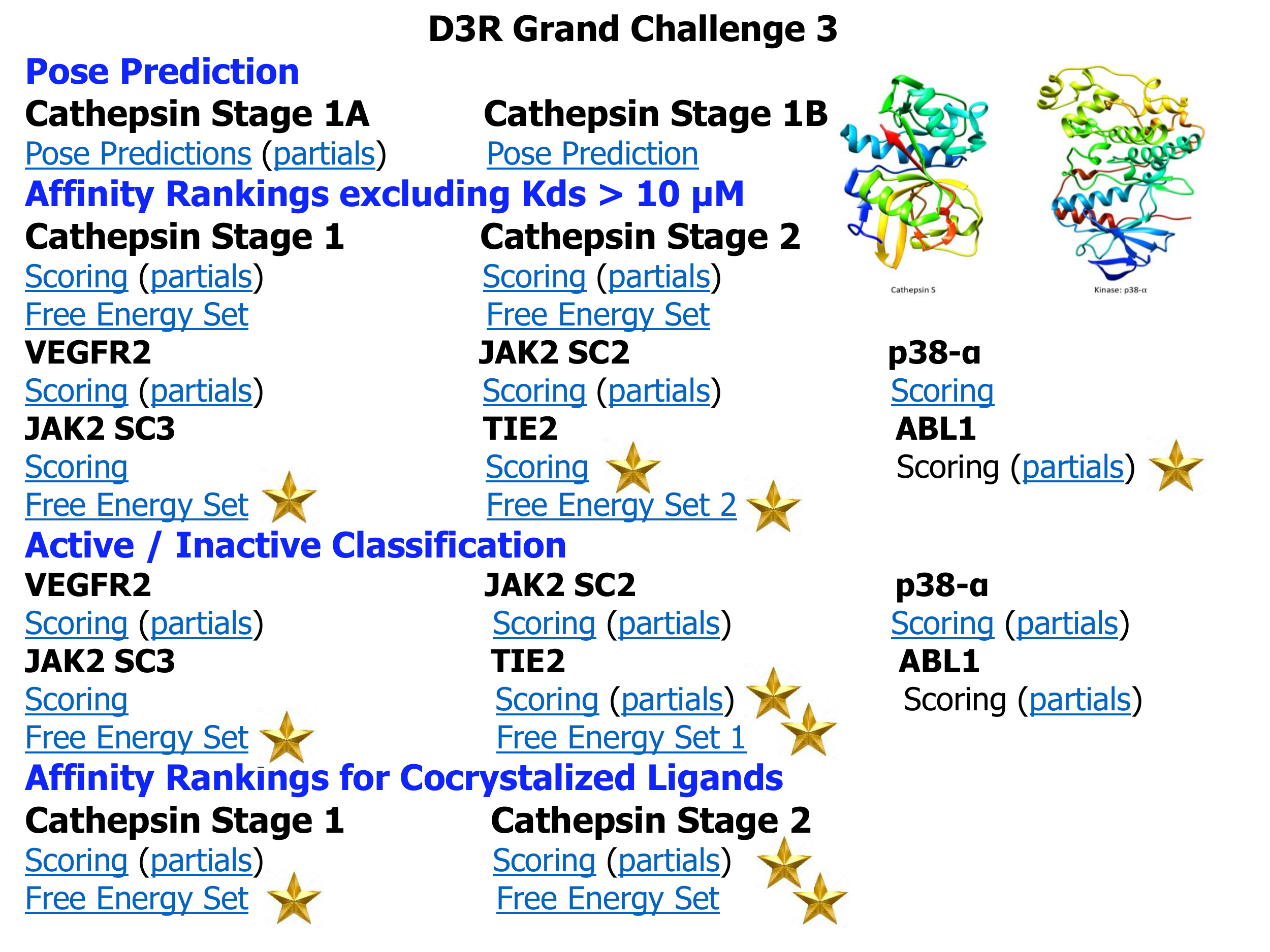}
	\caption{Overview of all 26 predictive tasks in D3R GC3. Our predictions were ranked 1st in the tasks marked by golden stars.}
	\label{fig:GC3_results}
\end{figure}
Here, we provide the results of our mathematical deep learning models in two current grand challenges, GC2 and GC3.
\subsection{Grand Challenge 3}
There are 5 subchallenges in GC3 involving a total of 12 affinity prediction submissions and 2 pose prediction challenges,  resulting in 26 different competitive tasks. Our submissions were ranked 1st  in 10 of these 26 tasks as shown in Fig. \ref{fig:GC3_results} for additional information.  While we employed align-close, align-target, close-dock and cross-dock protocols for pose generations in subchallenges 1--4, we applied constraint-IFD and free-IFD procedures for kinase mutants in subchallenge 5. The  combination of MWCG and algebraic topological descriptors was utilized as the features in the random forest and deep learning methods. Also, we were interested in seeing how the docking features can enhance our mathematical descriptors by including the Autodock Vina scoring terms in some submissions. In fact, these additional docking features did not improve our available models. The following is the discussion of our performance for each subchallenge task.

\paragraph{Subchallenge 1}
The protein target for this challenge is Cathepsin S. There are 24 ligand-protein co-crystal structures and 136 ligands having binding data (IC50s). There are two stages in this subchallenge. Stage 1 asks participants to submit pose prediction, affinity ranking, and energy prediction. Stage 2 asks similar tasks except for pose prediction. Co-crystal structures were released for the second stage.

\begin{figure}[!ht]
	\centering
	\includegraphics[width=\textwidth]{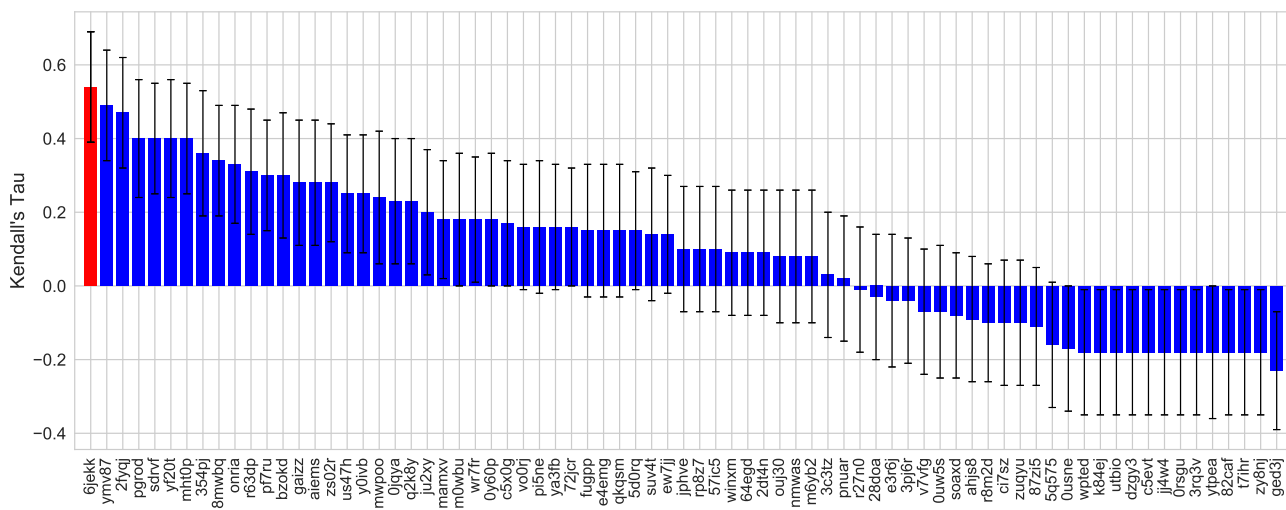}
	\caption{Performance comparison of different submissions on affinity ranking of 19 ligands having crystallographic poses in Stage 2 of Subchallenge 1 of D3R GC3. Our best prediction, with receipt ID 6jekk and in red color, achieved the top performance with Kendall's $\tau$ = 0.54.}
	\label{fig:CatS_stage2_ranking}
\end{figure}

In order to examine the performances of scoring functions on the binding affinity when the ligand pose errors do not contribute to the final outcome, D3R organizers evaluated the accuracy of all submitted methods on 19 ligands having crystallographic poses. With this setting, our models attained the first places for the following tasks: free energy set in Stage 1, scoring and free energy set in Stage 2. It is worth mentioning that only Stage 2 has the experimental structures. Stage 1 is still affected by the pose prediction errors. That explains why our predictors performed decently for scoring task in Stage 1 with the best Kendall's $\tau$ = 0.23, but they achieved a state-of-the-art result for the same task in Stage 2 with the best Kendall's $\tau$ = 0.54 (receipt ID 6jekk). Figure \ref{fig:CatS_stage2_ranking} depicts  the ranking of all participants on the affinity ranking of 19 ligands in Stage 2. The best free energy predictions on the ligands with experiment structures were also attained by our predictions. Particularly, in Stage 1, our prediction with receipt ID fomca obtained RMSE$_{\rm c}$=0.33 kcal/mol. In Stage 2, we accomplished RMSE$_{\rm c}$=0.29 kcal/mol with receipt ID v4jv4. Those results firmly support that our mathematical deep learning models indeed gain a better performance when no pose error predictions are involved.

\paragraph{Subchallenge 2}
In this subchallenge, there are 3 kinase families, namely VEGFGR2, JAK2, and p38-$\alpha$ with number of ligands being 85, 89 and 72, respectively. The challenge is to rank affinities of all ligands in each kinase family. Our predictors do not perform well on these datasets. Our best result is the second place on the active/inactive classification of VERGFR2 set. Our best Matthews correlation coefficient (MCC) on such task is reported to be 0.48 from receipt IDs: qikvs and rtv8m.

\paragraph{Subchallenge 3}
The third subchallenge involves the kinase JAK2 which already appeared in the second one. However, this challenge only comprises 17 compounds with small changes in chemical structure. Subchallenge 3 consists of two tasks, namely affinity ranking and relative binding affinity prediction. We obtained the first place on the binding energy prediction with the centered RMSE as low as RMSE$_{\rm c}$=1.06 kcal/mol (receipt ID 4u5ey). On the affinity ranking, the performance of our models is not impressive. However, we still manage to sit at the second place on the active/inactive classification with Mathew correlation coefficient = 0.23 with receipt ID yqoad.

\paragraph{Subchallenge 4}
Similar to the third subchallenge, the fourth one consists of 18 ligands with small changes in chemical structure. However, the new protein family, TIE2, is considered. The tasks are still to give an affinity ranking for 18 ligands and absolute or relative binding energies for two subsets of 4 and 6 compounds. It is interesting to see our model perform extremely well for the TIE2 dataset. We achieve the first place for all the evaluation metrics taken into account for this subchallenge. Specifically, for the affinity ranking excluding Kds > 10 $\mu{\rm M}$, our model, receipt ID uuihe, produces the best Kendall's $\tau$ and Spearman correlation coefficient among all of the  participants with values being 0.57 and 0.76, respectively. When one is interested in active/inactive classification by including compounds havingKds > 10 $\mu{\rm M}$, our model, receipt ID uuihe, is still ranked the first place with MCC = 0.78. On the absolute free energy predictions, the top results are still produced by our models. Specifically, on Set 1, our predictor with receipt ID vwbp8 was ranked the first place with MCC = 1.0. On Set 2, our model with receipt ID 5g8ed attained the RMSE$_{\rm c}$=1.02 kcal/mol which is the lowest among all submissions.

\paragraph{Subchallenge 5}
The last subchallenge in the GC3  measures the accuracy of models on the binding affinity change prediction upon the mutation. ABL1 is the protein target for this subchallenge, and there are two compounds and five mutants. The challenge is to predict the ranking of all mutants for each of two ligands. Our models perform pretty decently for this task. Our best submission has receipt ID rdn3k which achieves the best Kendall's tau ($\tau$=0.52) for affinity ranking excluding Kds > 10$\mu{\rm M}$.

\subsection{Grand challenge 2} The second grand challenge had 36 ligands with crystal structures and binding data for 102 ligands. All these compounds bind to the FXR target. The predictive tasks are the same as those of Subchallenge 1 in GC3. Specifically, GC2 consisted of two stages. The first stage  included (i) pose prediction for 36 ligands; (ii) binding affinity ranking for 102 compounds; (iii) absolute or relative free energy predictions for two subsets of 18 and 15 ligands, respectively. The second stage with released structures asked the same tasks as in the previous one except for pose prediction.

\begin{figure}[!ht]
	\centering
	\includegraphics[width=\textwidth]{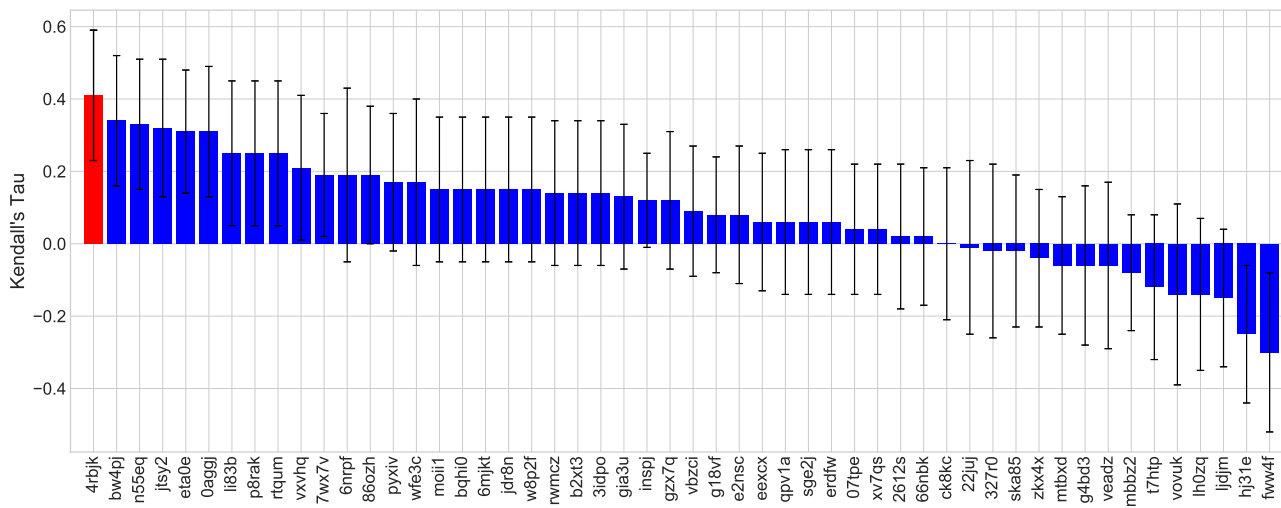}
	\caption{Performance comparison of different submissions on free energy prediction for free energy Set 1 in Stage 2 of D3R GC2. Our best prediction, with receipt ID 4rbjk and in red color, achieved the top performance with Kendall's $\tau$ = 0.41.}
	\label{fig:GC2_stage2_FE1}
\end{figure}

We employed the machine learning based scoring function to select the best poses for all prediction tasks. Although our pose ranking power was not impressive, the free energy predictions of our model performed pretty well. Specifically, our submission with receipt ID 5bvwx was ranked the second place in the free energy Set 1 of Stage 1 with RMSE$_{\rm c}$ = 0.68 kcal/mol. In Stage 2, our models improved the accuracy of the energy prediction of compounds in the aforementioned free energy set. In fact, we obtained the first place in term of Kendall's tau value ($\tau=0.41$) with receipt ID 4rbjk. That was also the highest Kendall's tau value among all submissions in two stages for the free energy Set 1. Figure \ref{fig:GC2_stage2_FE1} plots the performance of all submissions on the free energy Set 1 in Stage 2. Our best submission is highlighted in the red color.

\section{Conclusion}
In this work, we report the performances of our mathematical deep learning strategy on the binding affinity tasks in D3R GC2 and across 5 subchallenges in D3R GC3. The multiscale weighted colored graph and element specific persistent homology  representations are the main descriptors in our models. We also employed a variety of machine learning algorithms including random forest and deep convolutional neural networks for the energy predictions. Overall, in GC2, our predictive models achieved the top place in free energy prediction for free energy Set 1 in Stage 2. In GC3, our submissions were ranked 1st in 10 out of 26 official evaluation tasks. These results rigorously confirm the predictive power and practical usage  of our mathematical deep learning models in drug discovery. It is worthy to mention that the docking accuracy is still a bottleneck of our affinity prediction performance. We have tried a variety of docking protocols, namely align-close, align-target, close-dock, cross-dock, constraint-IFD, and free-IFD, for pose selection in GC3. However, none of them showed a dominant role in binding affinity accuracy.
In addition, when one excludes the pose prediction error, Kendall's tau of our model improves from 0.21 to 0.54 on the affinity ranking of compounds in Cathepsin S subchallenge. Therefore, the development of a state-of-the-art docking protocol is our major task in the roadmap to improve the accuracy of binding energy prediction when crystallographic structures are not available.

 \section*{Acknowledgments}

This work was supported in part by NSF grants  IIS-1302285, DMS-1160352 and DMS-1761320 and
MSU Center for Mathematical Molecular Biosciences Initiative.

%

\begin{thebibliography}{10}

\bibitem{Berman:2000}
H.~M. Berman, J.~Westbrook, Z.~Feng, G.~Gilliland, T.~N. Bhat, H.~Weissig,
  I.~N. Shindyalov, and P.~E. Bourne, ``The protein data bank,'' {\em Nucleic
  acids research}, vol.~28, no.~1, pp.~35--242, 2000.

\bibitem{liu2017forging}
Z.~Liu, M.~Su, L.~Han, J.~Liu, Q.~Yang, Y.~Li, and R.~Wang, ``Forging the basis
  for developing protein--ligand interaction scoring functions,'' {\em Accounts
  of Chemical Research}, vol.~50, no.~2, pp.~302--309, 2017.

\bibitem{ahmed2014recent}
A.~Ahmed, R.~D. Smith, J.~J. Clark, J.~B. Dunbar~Jr, and H.~A. Carlson,
  ``Recent improvements to binding moad: a resource for protein--ligand binding
  affinities and structures,'' {\em Nucleic acids research}, vol.~43, no.~D1,
  pp.~D465--D469, 2014.

\bibitem{kroemer2007structure}
R.~T. Kroemer, ``Structure-based drug design: docking and scoring,'' {\em
  Current protein and peptide science}, vol.~8, no.~4, pp.~312--328, 2007.

\bibitem{Leach:2006}
A.~R. Leach, B.~K. Shoichet, and C.~E. Peishoff, ``Prediction of protein-ligand
  interactions. docking and scoring: Successes and gaps.,'' {\em J. Med.
  Chem.}, vol.~49, pp.~5851--5855, 2006.

\bibitem{Novikov:2011}
F.~N. Novikov, A.~A. Zeifman, O.~V. Stroganov, V.~S. Stroylov, V.~Kulkov, and
  G.~G. Chilov, ``{CSAR Scoring} challenge reveals the need for new concepts in
  estimating protein-ligand binding affinity,'' {\em Journal of Chemical
  Information and Model}, vol.~51, pp.~2090--2096, 2011.

\bibitem{RenxiaoWang:2003CompareSF}
R.~Wang, Y.~Lu, and S.~Wang, ``Comparative evaluation of 11 scoring functions
  for molecular docking,'' {\em J. Med. Chem.}, vol.~46, pp.~2287--2303, 2003.

\bibitem{LiuJie:2014}
J.~Liu and R.~Wang, ``Classification of current scoring functions,'' {\em
  Journal of Chemical Information and Model}, vol.~55, no.~3, pp.~475--482,
  2015.

\bibitem{Ortiz:1995}
A.~R. Ortiz, M.~T. Pisabarro, F.~Gago, and R.~C. Wade, ``Prediction of drug
  binding affinities by comparative binding energy analysis,'' {\em J. Med.
  Chem}, vol.~38, pp.~2681--2691, 1995.

\bibitem{Yin:2008}
S.~Yin, L.~Biedermannova, J.~Vondrasek, and N.~V. Dokholyan, ``Medusascore: An
  acurate force field-based scoring function for virtual drug screening,'' {\em
  Journal of Chemical Information and Model}, vol.~48, pp.~1656--1662, 2008.

\bibitem{Merz:2015Solvation}
Z.~Zheng, T.~Wang, P.~Li, and K.~M. Merz~Jr, ``{KECSA-Movable} type implicit
  solvation model {(KMTISM)},'' {\em Journal of Chemical Theory and
  Computation}, vol.~11, pp.~667--682, 2015.

\bibitem{PMFScore:1999}
I.~Muegge and Y.~Martin, ``A general and fast scoring function for
  protein-ligand interactions: a simplified potential approach.,'' {\em J Med
  Chem.}, vol.~42, no.~5, pp.~791--804, 1999.

\bibitem{DrugScore:2005}
H.~F.~G. Velec, H.~Gohlke, and G.~Klebe, ``Knowledge-based scoring function
  derived from small molecule crystal data with superior recognition rate of
  near-native ligand poses and better affinity prediction.,'' {\em J. Med.
  Chem}, vol.~48, pp.~6296--6303, 2005.

\bibitem{ITScore:2006}
S.~Y. Huang and X.~Zou, ``An iterative knowledge-based scoring function to
  predict protein-ligand interactions: I. derivation of interaction
  potentials.,'' {\em J. Comput. Chem.}, vol.~27, pp.~1865--1875, 2006.

\bibitem{XScore:2002}
R.~Wang, L.~Lai, and S.~Wang, ``Further development and validation of empirical
  scoring functions for structural based binding affinity prediction.,'' {\em
  J. Comput-Aided Mol. Des}, vol.~16, pp.~11--26, 2002.

\bibitem{Verkhivker:1995PLP}
G.~Verkhivker, K.~Appelt, S.~T. Freer, and J.~E. Villafranca, ``Empirical free
  energy calculations of ligand-protein crystallographic complexes. i.
  knowledge based ligand-protein interaction potentials applied to the
  prediction of human immunodeficiency virus protease binding affinity.,'' {\em
  Protein Eng}, vol.~8, pp.~677--691, 1995.

\bibitem{Eldridge:1997}
M.~D. Eldridge, C.~W. Murray, T.~R. Auton, G.~V. Paolini, and R.~P. Mee,
  ``Empirical scoring functions: I. the development of a fast empirical scoring
  function to estimate the binding affinity of ligands in receptor
  complexes.,'' {\em J. Comput. Aided. Mol. Des}, vol.~11, pp.~425--445, 1997.

\bibitem{Baum:2010}
B.~Baum, L.~Muley, M.~Smolinski, A.~Heine, D.~Hangauer, and G.~Klebe,
  ``Non-additivity of functional group contributions in protein-ligand binding:
  a comprehensive study by crystallography and isothermal titration
  calorimetry,'' {\em J. Mol. Bio}, vol.~397, no.~4, pp.~1042--1054, 2010.

\bibitem{li:2014}
H.~Li, K.-S. Leung, M.-H. Wong, and P.~J. Ballester, ``Substituting random
  forest for multiple linear regression improves binding affinity prediction of
  scoring functions: {Cyscore} as a case study,'' {\em BMC bioinformatics},
  vol.~15, no.~1, p.~1, 2014.

\bibitem{DDNguyen:2017d}
D.~D. Nguyen, T.~Xiao, M.~L. Wang, and G.~W. Wei, ``{ Rigidity strengthening: A
  mechanism for protein-ligand binding },'' {\em Journal of Chemical
  Information and Modeling}, vol.~57, pp.~1715--1721, 2017.

\bibitem{ZXCang:2017b}
Z.~X. Cang and G.~W. Wei, ``{Integration of element specific persistent
  homology and machine learning for protein-ligand binding affinity prediction
  },'' {\em International Journal for Numerical Methods in Biomedical
  Engineering}, vol.~34(2), p.~DOI: 10.1002/cnm.2914, 2018.

\bibitem{ZXCang:2017c}
Z.~X. Cang and G.~W. Wei, ``{TopologyNet: Topology based deep convolutional and
  multi-task neural networks for biomolecular property predictions},'' {\em
  PLOS Computational Biology}, vol.~13(7), pp.~e1005690,
  https://doi.org/10.1371/journal.pcbi.1005690, 2017.

\bibitem{ZXCang:2018a}
Z.~X. Cang, L.~Mu, and G.~W. Wei, ``{Representability of algebraic topology for
  biomolecules in machine learning based scoring and virtual screening },''
  {\em PLOS Computational Biology}, vol.~14(1), pp.~e1005929,
  https://doi.org/10.1371/journal.pcbi.1005929, 2018.

\bibitem{bramer2018multiscale}
D.~Bramer and G.-W. Wei, ``Multiscale weighted colored graphs for protein
  flexibility and rigidity analysis,'' {\em The Journal of chemical physics},
  vol.~148, no.~5, p.~054103, 2018.

\bibitem{Kaczynski:2004}
T.~Kaczynski, K.~Mischaikow, and M.~Mrozek, {\em Computational homology}.
\newblock Springer-Verlag, 2004.

\bibitem{Edelsbrunner01topologicalpersistence}
H.~Edelsbrunner, D.~Letscher, and A.~Zomorodian, ``Topological persistence and
  simplification,'' {\em Discrete Comput. Geom}, vol.~28, pp.~511--533, 2001.

\bibitem{Zomorodian:2005}
A.~Zomorodian and G.~Carlsson, ``Computing persistent homology,'' {\em Discrete
  Comput. Geom.}, vol.~33, pp.~249--274, 2005.

\bibitem{Frosini:1999}
P.~Frosini and C.~Landi, ``Size theory as a topological tool for computer
  vision,'' {\em Pattern Recognition and Image Analysis}, vol.~9, no.~4,
  pp.~596--603, 1999.

\bibitem{Kasson:2007}
P.~M. Kasson, A.~Zomorodian, S.~Park, N.~Singhal, L.~J. Guibas, and V.~S.
  Pande, ``Persistent voids a new structural metric for membrane fusion,'' {\em
  Bioinformatics}, vol.~23, pp.~1753--1759, 2007.

\bibitem{Gameiro:2014}
M.~Gameiro, Y.~Hiraoka, S.~Izumi, M.~Kramar, K.~Mischaikow, and V.~Nanda,
  ``Topological measurement of protein compressibility via persistence
  diagrams,'' {\em Japan Journal of Industrial and Applied Mathematics},
  vol.~32, pp.~1--17, 2014.

\bibitem{dabaghian2012topological}
Y.~Dabaghian, F.~M{\'e}moli, L.~Frank, and G.~Carlsson, ``A topological
  paradigm for hippocampal spatial map formation using persistent homology,''
  {\em PLoS computational biology}, vol.~8, no.~8, p.~e1002581, 2012.

\bibitem{KLXia:2014c}
K.~L. Xia and G.~W. Wei, ``Persistent homology analysis of protein structure,
  flexibility and folding,'' {\em International Journal for Numerical Methods
  in Biomedical Engineering}, vol.~30, pp.~814--844, 2014.

\bibitem{KLXia:2015a}
K.~L. Xia, X.~Feng, Y.~Y. Tong, and G.~W. Wei, ``Persistent homology for the
  quantitative prediction of fullerene stability,'' {\em Journal of
  Computational Chemistry}, vol.~36, pp.~408--422, 2015.

\bibitem{BaoWang:2016a}
B.~Wang and G.~W. Wei, ``Object-oriented persistent homology,'' {\em Journal of
  Computational Physics}, vol.~305, pp.~276--299, 2016.

\bibitem{ESES:2017}
B.~Liu, B.~Wang, R.~Zhao, Y.~Tong, and G.~W. Wei, ``{ESES: software for
  Eulerian solvent excluded surface},'' {\em Journal of Computational
  Chemistry}, vol.~38, pp.~446--466, 2017.

\bibitem{KLXia:2015b}
K.~L. Xia and G.~W. Wei, ``Persistent topology for {cryo-EM} data analysis,''
  {\em International Journal for Numerical Methods in Biomedical Engineering},
  vol.~31, p.~e02719, 2015.

\bibitem{ZXCang:2015}
Z.~X. Cang, L.~Mu, K.~Wu, K.~Opron, K.~Xia, and G.-W. Wei, ``A topological
  approach to protein classification,'' {\em Molecular based Mathematical
  Biology}, vol.~3, pp.~140--162, 2015.

\bibitem{ZXCang:2017a}
Z.~X. Cang and G.~W. Wei, ``{Analysis and prediction of protein folding energy
  changes upon mutation by element specific persistent homology},'' {\em
  Bioinformatics}, vol.~33, pp.~3549--3557, 2017.

\bibitem{KDWu:2018a}
K.~Wu and G.-W. Wei, ``{Quantitative Toxicity Prediction Using Topology Based
  Multitask Deep Neural Networks},'' {\em Journal of Chemical Information and
  Modeling}, p.~http://dx.doi.org/10.1021/acs.jcim.7b00558, 2018.

\bibitem{wu2017topp}
K.~Wu, Z.~Zhao, R.~Wang, and G.-W. Wei, ``Topp-s: Persistent homology based
  multi-task deep neural networks for simultaneous predictions of partition
  coefficient and aqueous solubility,'' {\em arXiv preprint arXiv:1801.01558},
  2017.

\bibitem{Madhavi:2013Schrodinger}
G.~M. Sastry, M.~Adzhigirey, T.~Day, R.~Annabhimoju, and W.~Sherman, ``Protein
  and ligand preparation: parameters, protocols, and influence on virtual
  screening enrichments.,'' {\em J. Comput. Aid. Mol. Des.}, vol.~27,
  pp.~221--234, 2013.

\bibitem{Trott:2010AutoDock}
O.~Trott and A.~J. Olson, ``{AutoDock Vina}: improving the speed and accuracy
  of docking with a new scoring function, efficient optimization, and
  multithreading,'' {\em J Computat Chem}, vol.~31, no.~2, pp.~455--461, 2010.

\bibitem{morris2009autodock4}
G.~M. Morris, R.~Huey, W.~Lindstrom, M.~F. Sanner, R.~K. Belew, D.~S. Goodsell,
  and A.~J. Olson, ``Autodock4 and autodocktools4: Automated docking with
  selective receptor flexibility,'' {\em Journal of computational chemistry},
  vol.~30, no.~16, pp.~2785--2791, 2009.

\bibitem{bell2012primex}
J.~Bell, Y.~Cao, J.~Gunn, T.~Day, E.~Gallicchio, Z.~Zhou, R.~Levy, and
  R.~Farid, ``Primex and the schr{\"o}dinger computational chemistry suite of
  programs,'' 2012.

\bibitem{ye2016optimal}
Z.~Ye, M.~P. Baumgartner, B.~M. Wingert, and C.~J. Camacho, ``Optimal
  strategies for virtual screening of induced-fit and flexible target in the
  2015 d3r grand challenge,'' {\em Journal of computer-aided molecular design},
  vol.~30, no.~9, pp.~695--706, 2016.

\bibitem{G-Score}
G.~Jones, P.~Willett, R.~C. Glen, A.~R. Leach, and R.~Taylor, ``Development and
  validation of a genetic algorithm for flexible docking.,'' {\em Journal of
  Molecular Biology}, vol.~267, no.~3, pp.~727--748, 1997.

\bibitem{Friesner:2004Schrodinger}
R.~A. Friesner, J.~L. Banks, R.~B. Murphy, T.~A. Halgren, J.~J. Klicic, D.~T.
  Mainz, M.~P. Repasky, E.~H. Knoll, M.~Shelley, J.~K.~P. JK, D.~E. Shaw,
  P.~Francis, and P.~S. Shenkin, ``Glide: a new approach for rapid, accurate
  docking and scoring. 1. method and assessment of docking accuracy.,'' {\em J.
  Med. Chem.}, vol.~47, p.~1739, 2004.

\bibitem{o:2011}
N.~M. O'Boyle, M.~Banck, C.~A. James, C.~Morley, T.~Vandermeersch, and G.~R.
  Hutchison, ``Open babel: An open chemical toolbox,'' {\em Journal of
  cheminformatics}, vol.~3, no.~1, p.~1, 2011.

\bibitem{schrodinger2017-4}
S.~LLC, ``Schr\"{o}dinger release 2017-4, {Schr\"{o}dinger LLC}, new york,''
  2017.

\bibitem{dixon2006phase}
S.~L. Dixon, A.~M. Smondyrev, E.~H. Knoll, S.~N. Rao, D.~E. Shaw, and R.~A.
  Friesner, ``Phase: a new engine for pharmacophore perception, 3d qsar model
  development, and 3d database screening: 1. methodology and preliminary
  results,'' {\em Journal of computer-aided molecular design}, vol.~20,
  no.~10-11, pp.~647--671, 2006.

\bibitem{dixon2006phase2}
S.~L. Dixon, A.~M. Smondyrev, and S.~N. Rao, ``Phase: a novel approach to
  pharmacophore modeling and 3d database searching,'' {\em Chemical biology \&
  drug design}, vol.~67, no.~5, pp.~370--372, 2006.

\bibitem{jacobson2004hierarchical}
M.~P. Jacobson, D.~L. Pincus, C.~S. Rapp, T.~J. Day, B.~Honig, D.~E. Shaw, and
  R.~A. Friesner, ``A hierarchical approach to all-atom protein loop
  prediction,'' {\em Proteins: Structure, Function, and Bioinformatics},
  vol.~55, no.~2, pp.~351--367, 2004.

\bibitem{jacobson2002role}
M.~P. Jacobson, R.~A. Friesner, Z.~Xiang, and B.~Honig, ``On the role of the
  crystal environment in determining protein side-chain conformations,'' {\em
  Journal of molecular biology}, vol.~320, no.~3, pp.~597--608, 2002.

\bibitem{farid2006new}
R.~Farid, T.~Day, R.~A. Friesner, and R.~A. Pearlstein, ``New insights about
  herg blockade obtained from protein modeling, potential energy mapping, and
  docking studies,'' {\em Bioorganic \& medicinal chemistry}, vol.~14, no.~9,
  pp.~3160--3173, 2006.

\bibitem{sherman2006novel}
W.~Sherman, T.~Day, M.~P. Jacobson, R.~A. Friesner, and R.~Farid, ``Novel
  procedure for modeling ligand/receptor induced fit effects,'' {\em Journal of
  medicinal chemistry}, vol.~49, no.~2, pp.~534--553, 2006.

\bibitem{sherman2006use}
W.~Sherman, H.~S. Beard, and R.~Farid, ``Use of an induced fit receptor
  structure in virtual screening,'' {\em Chemical biology \& drug design},
  vol.~67, no.~1, pp.~83--84, 2006.

\bibitem{borgatti2005centrality}
S.~P. Borgatti, ``Centrality and network flow,'' {\em Social networks},
  vol.~27, no.~1, pp.~55--71, 2005.

\bibitem{freeman1978centrality}
L.~C. Freeman, ``Centrality in social networks conceptual clarification,'' {\em
  Social networks}, vol.~1, no.~3, pp.~215--239, 1978.

\bibitem{bavelas1950communication}
A.~Bavelas, ``Communication patterns in task-oriented groups,'' {\em The
  Journal of the Acoustical Society of America}, vol.~22, no.~6, pp.~725--730,
  1950.

\bibitem{dekker2005conceptual}
A.~Dekker, ``Conceptual distance in social network analysis,'' {\em Journal of
  Social Structure (JOSS)}, vol.~6, 2005.

\bibitem{Alpha}
H.~Edelsbrunner, ``Weighted alpha shapes,'' tech. rep., Champaign, IL, USA,
  1992.

\bibitem{DDNguyen:2018b}
D.~D. Nguyen and G.~W. Wei, ``Multiscale weighted colored algebraic graphs for
  biomolecules,'' To be submitted 2018.

\end{thebibliography}

\end{document}